\documentclass[5p,times]{elsarticle}

\usepackage{graphicx}
\usepackage{lineno}
\usepackage{lscape}
\usepackage{tabularx}
\usepackage[gen]{eurosym}
\usepackage{multirow}
\usepackage{hhline}
\usepackage{booktabs}
\usepackage{siunitx}
\usepackage{framed}
\usepackage{float}
\usepackage{amssymb,amsmath} 
\usepackage[caption = false]{subfig}
\usepackage{siunitx}
\ExplSyntaxOn
\makeatletter
\AtBeginDocument
  {
    \@ifpackageloaded { mdwtab }
      {
        \cs_set_protected:Npn \__siunitx_table_collect_not_braced:N #1
          {
            \token_if_eq_meaning:NNF #1 \tex_ignorespaces:D
              {
                \token_if_eq_meaning:NNF #1 \tex_unskip:D
                  {
                    \token_if_eq_meaning:NNF #1 \tab@setcr
                      {
                        \token_if_eq_meaning:NNF #1 \@maybe@unskip
                          { \__siunitx_table_collect_not_braced_aux_i:N #1 }
                      }  
                  }
              }
            \__siunitx_table_collect_next:
          }
      }
      { }
  }
\makeatother
\ExplSyntaxOff

\usepackage{mdwtab}

\usepackage{url}

\usepackage{pdfpages}
\usepackage{csquotes}
\usepackage{array} % loaded by booktabs, but included here for explicitness
\usepackage{booktabs}
\usepackage{siunitx}

\newcolumntype{L}{@{}l@{}} % a left column with no intercolumn space on either side
\newcommand{\mc}[1]{\multicolumn{1}{c}{#1}} % shorthand macro for column headings

\usepackage[utf8]{inputenc}
\usepackage{changepage}% http://ctan.org/pkg/changepage
\usepackage{lipsum}% http://ctan.org/pkg/lipsum

\usepackage{amsthm}

\journal{IEEE Software: Special Issue on Sentiment and Emotion in Software Engineering}

\begin{document}

\begin{frontmatter}

\title{The Connection Between Burnout and Personality Types in Software Developers}

\author{Emanuel Mellblom}
\ead{gusmellbem@student.gu.se}

\author{Isar Arason}
\ead{gusarais@student.gu.se}

\author{Lucas Gren\corref{cor1}}
\ead{lucas.gren@cse.gu.se}

\author{Richard Torkar}
\ead{richard.torkar@cse.gu.se}

\address{The Department of Computer Science and Engineering, Chalmers University of Technology and the University of Gothenburg, Gothenburg, Sweden}

\begin{abstract}
This paper examines the connection between the Five Factor Model personality traits and burnout in software developers. This study aims to validate generalizations of findings in other fields. An online survey consisting of a miniaturized International Personality Item Pool questionnaire for measuring the Five Factor Model personality traits, and the Shirom-Melamed Burnout Measure for measuring burnout, were distributed to open source developer mailing lists, obtaining 47 valid responses. The results from a Bayesian Linear Regression analysis indicate a strong link between neuroticism and burnout confirming previous work, while the other Five Factor Model traits were not adding power to the model. It is important to note that we did not investigate the quality of work in connection to personality, nor did we take any other confounding factors into account like, for example, teamwork. Nonetheless, employers could be aware of, and support, software developers with high neuroticism. 
\end{abstract}
%150 words-- yes!


\end{frontmatter}
%\linenumbers

%Manuscripts must not exceed 3,000 words including figures and tables, which count for 250 words each. 

\section{Introduction}
Stress is a commonplace occurrence in many professions. It is a natural response to averse circumstances, heightening alertness and improving reaction speed, allowing a person to better handle a stressful situation. However, this state takes a toll on the human body, impacting the mental and physical health of a person if sustained for long periods of time \citep{burnouteffects}.

People which experience long-term stress on the job are at greater risk of burnout. Burnout occurs when a person can no longer handle the stress put upon them effectively, resulting in reduced work efficiency, unhappiness, and health problems. However, some people appear to thrive in high intensity environments for long periods of time without burnout \citep{burnouteffects}. This may be related to the apparent link between personality types and stress suggested in other studies (e.g.\ \citet{ffm_meta}). We would like to highlight already that personality is one aspect of a person and other aspects might even be more important to avoid stress, like for example social support \citep{cobb1976social}.

A form of personality measurement is the Five Factor Model (FFM) also called the Big Five. The FFM is a construct composed of five personality types: Openness to experience, Conscientiousness, Extraversion, Agreeableness, and Neuroticism \citep{bigfivemarkers}. We aim to investigate the connection between the FFM personality types and burnout in software engineers. This may provide some insight into why some people are more resistant to burnout than others, which in turn would give employers and potential employees an additional tool for managing workplace stress and burnout. We would suspect software developers to have similar connections between personality types and burnout as people in other professions, however, there might be differences we cannot predict. If that were the case, it could, either guide us in future studies on personality and burnout in software developers, or we might simply have too weak connections that imply we should look at other construct to help software developers avoid burnout. 

A number of different personality traits, derived from a multitude of different methods, have been correlated with burnout in other fields, e.g.\ \citet{bigfiveburnout}, but this paper appears to be first to investigate this apparent correlation in the context of software developers. 

{\textbf{$RQ$:}} {\em Are any of the Big Five personality traits associated with software developer burnout?}

\section{Background}

\subsection{Burnout and occupational stress}

The concept of burnout was originally formulated in the mid 70s. Since the concept was first coined, interest in this subject has grown significantly and the amount of research on burnout has increased considerably. The most used definition of burnout was created in 1982 by \citet{occupationalburnout}. This model has been modified over the years as research has found that burnout can occur in all professions. The model now covers all occupations whereas it was initially focused solely on health-care professionals. The three constructs of Maslach's burnout model (MBI) are emotional exhaustion, depersonalization, and personal accomplishment. Emotional exhaustion can occur when an employee experiences overextended periods of stress, often caused by work overload or conflicts. The second construct, depersonalization or cynicism, represents a negative response to other people, as well as a feeling of being disconnected from others. The last component, personal accomplishment, is described as a decrease in an individual’s feeling of competence and productivity \citep{occupationalburnout}.

A recent meta-study show that there is a 3 -- 15\%  prevalence of severe burnout across professions and little research in what interventions that actually are effective \citep{ahola2017interventions}. It should therefore be of high importance for employees and employers to take a preventive approach to burnout. On the personal plane, an employee or individual can engage in some person-centered approaches that have been showed to help prevent burnout. Some of these prevention methods includes engaging in relaxing activities such as yoga, meditation and mindfulness training. Another method is to take a vacation which allows the employee to take a break from the stressors at work. It has also been demonstrated that social support is one of the most efficient ways of preventing burnout, (e.g.\ \citep{associationsbetween}). Past studies have found that burnout decreases the efficiency and quality of work of employees \citep{occupationalburnout}.

\subsection{Big Five Model}
After many studies involving a diversity of traits, five main factors were selected among them through statistical factor analysis techniques. This later led to the release of the Big Five, also known as the Five Factor Model (FFM) \citep{bigfivemarkers}.

The five factor model consists of five main personality types, often abbreviated OCEAN, each with different associated facets and are as follows:
  \paragraph {Openness to Experience (O)} People with high scores in Openness to Experience are often related to creativity both in artistic ways as well as in a scientific way. They are described as having divergent thinking and low religiosity, as well as liberal in politics. A high score in this category implies one likes to learn new things, is imaginative, has a variety of interests, and finds enjoyment in new experiences.
  \paragraph {Conscientiousness (C)} Conscientious individuals are often career oriented. They also tend to have a high job satisfaction. Individuals scoring high in this category are described as organized, thorough, and methodical.
  \paragraph {Extraversion (E)} Extraverts enjoy engaging in social interactions and have more friends. An extraverted person often feels energetic and talkative. People in this category often get their energy and drive from other people.
  \paragraph {Agreeableness (A)} People high on Agreeableness often experience happiness and high life satisfaction since they are less critical, kind, sympathetic, cooperative, warm, and considerate. They also find it easier to engage in relationships with others. 
  \paragraph {Neuroticism (N)} Neurotic individuals are characterized as being more susceptible to negative emotions, and experience more feelings of anger, worry, envy, fear, anxiety, jealousy, guilt, and frustration. Individuals scoring high in neuroticism are more likely to describe any given event as a negative experience and have a tendency to describe situations as threatening. This in turn can lead to decreased job satisfaction and neurotic individuals are classified as being susceptible to negative emotions and emotional instability \citep{persjob}.

\subsection{SMBQ\slash SMBM}
As a burnout measurement, we used the Shirom Melamed Burnout Measure (SMBM), which is a burnout measure derived from Shirom-Melamed Burnout Questionnaire (SMBQ). Recent studies indicate that the MBI mentioned above and SMBM essentially measure the same thing \citep{jocic2018cultural} but we chose SMBM due to its accessibility and shortness \citep{burnoutmeasure}. Burnout in SMBQ is defined as a construct that consists of emotional exhaustion, physical fatigue, and cognitive weariness. These components together represent burnout. This measure is also a widely used method for measuring burnout with high validity \citep{constructvalidity}.

\subsection{IPIP}
The International Personality Item Pool (IPIP) is a collection of items used to measure a variety of personality related constructs, including the Five Factor Model. Compared to other measures, IPIP has been found to be more consistent in describing FFM personality traits. The measure can be conducted in the form of a questionnaire consisting of factors and sub-factors which provide numerical values, enabling detailed statistical analysis \citep{linkssoftwaredev}. IPIP is public and free to copy, edit, and use without explicit permission or fees.

\subsection{Previous research on personality and burnout}
Previous research has found indications of personality being an influencing factor in the development of burnout. High perceived job strain is primarily associated with people scoring high in Neuroticism, whereas people scoring high in Extraversion, Openness, Conscientiousness, and Agreeableness perceived their work to be less stressful, and generally experienced lower job strain \citep{associationsbetween}. Studies have found that introverts often get stressed more easily than their counterparts \citep{grant2007personality}. Furthermore, Neuroticism and Conscientiousness scores may be used as predictors of burnout \citep{bigfivefactorspredicators}. People with high scores in Conscientiousness appear to be less likely to experience burnout, but the personality type most often found to be associated with stress, burnout, and strain is Neuroticism \citep{associationsbetween}.

\subsection{Other confounding factors}
It is very important to note that personality is only one factor in explaining burnout and e.g.\ work-family conflict \citep{leineweber2012work}, decision latitude at work \citep{karasek1979job}, and social support \citep{cobb1976social} have all been shown to be very important factors in relation to stress and burnout and were not a part of this study. %team?

\section{Method}
For measuring personality traits, we used a miniaturized IPIP questionnaire \citep{miniipip}. This miniaturized version has been validated as closely matching the original 50 item IPIP FFM questionnaire, which has been found to better describe personality than other available personality tests, such as MBTI. Its smaller size also makes it better suited for use in an online format where respondents may have limited time or patience for answering a survey, and IPIP is also freely available \citep{linkssoftwaredev}. This questionnaire outputs five variables, one for each personality trait, with values ranging from 1 to 5, with 1 meaning very inaccurate, and 5 meaning very accurate. 

For measuring burnout levels, SMBM was used. SMBM is very short, consisting of only 14 questions, making it ideal for use in a voluntary survey. SMBM is also specifically tailored to measure burnout in working populations, meeting our requirements for this study \citep{constructvalidity}. Each question in SMBM has answers ranging from 1 to 7, 1 representing never/almost never, and 7 representing always/almost always. The burnout coefficient was derived as the median of each answer in the questionnaire. 

The questionnaire consisted of three parts: an introduction and validation section, the IPIP section, and the SMBM section. The introduction contained a message describing the survey and the goal of the study. The validation section consisted of a yes or no question asking whether the participant was currently employed as a software engineer, followed by a question about years of experience. The former question is required as SMBM assumes ongoing employment. We had intended to take years of experience into account during the analysis, but this question was left unused post-collection as the experience groups were largely in the 10+ year range, leaving the remaining experience ranges too small for meaningful statistical analysis.

The data was collected through an online survey with 51 responses in total. This was done by distributing it through 17 open source mailing lists. The process that we followed for selecting which mailing lists to include in this study was as follows, first, we started to look for active open source projects using Google and searching for ``open source organizations.'' From the search result we created a list of organizations such as Mozilla, Apache, Redhat, KDE, Eclipse. From this list of organizations we started to look for mailing lists for these projects on their respective websites. The mailing lists that we decided to use were the ones specifically for software developers, as stated in the description for the mailing lists, resulting in the list of 17 mailing lists to include in this study. Each mailing lists was then joined by the authors and a mail with information about the study and the questionnaire was then submitted to each of the mailing lists. We also did an advertisement through Twitter from a Swedish software developer podcast named ``Kodsnack.'' The mailing list were to project with software developers from all over the world, which provided the potential av being representative for software developers overall. Four out of the 51 responses were invalidated due to responding negatively to being employed at the time of taking the questionnaire, which is required by the SMBM part of the survey. We do not know how many were on this list but the responses come from a fraction of the members on those open source mailing lists.

To extract data points from the survey results, each of the two sections of the questionnaire were processed following instructions outlined by their respective authors. In the IPIP part of the questionnaire, each item is accompanied by answers ranging from 1 to 5, where 1 means the subject strongly disagrees with the statement, and 5 means the subject strongly agrees. Each question is associated with a specific personality trait, either positively or negatively, and the value of the answer determines its weight when calculating the final score for each trait. Since we have ordinal data we used the median of these score to get an overall score on each trait. Similarly, the SMBM part's numerical answers were also calculated as a median for the final coefficient.

The null hypothesis ($H_{0}$) in a Bayesian linear regression analysis states that there is no added predictive power between a person's personality trait (Openness, Conscientiousness, Extraversion, Agreeableness, and Neuroticism) and their susceptibility to burnout. We tested all the different combinations of the five personality traits against a null model. We compared all the different models to the null model by using $BF_{10}$. For $BF_{10}$, the more then 30 is very strong evidence and the model with the highest value outperforms the other tested models.

\subsection{Validity Threats}
One threat, and arguably the most serious one, is selection bias. The sampled population is self-selected through voluntary participation in an online survey. This surveying method was chosen over other methods, such as in-person surveys, as it would allow the collection of more data points. An added drawback to online surveys is the lack of response rate measurement due to the distribution method of the survey. No information regarding the of number of people subscribed to each mailing list was available.

People participating in self-assessment surveys on personality have the tendency to not be fully honest with their answers, but studies have demonstrated that the IPIP framework is resistant to these effects \citep{linkssoftwaredev}. No such studies have been made regarding SMBM, which remains an uncertainty. However, this survey is done anonymously online mitigating this threat.

\section{Results}
\begin{figure*}
\subfloat[Openness to Experience.]{\includegraphics[width = 2in]{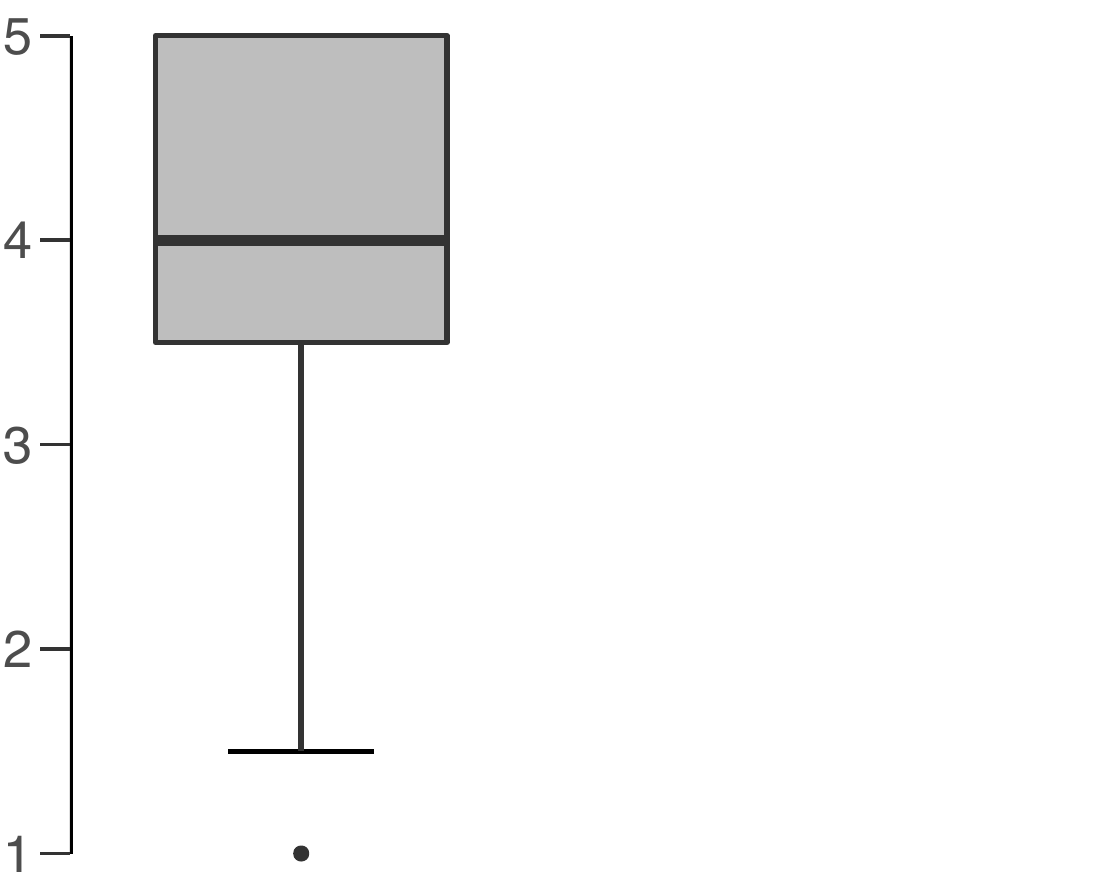}} 
\subfloat[Conscientiousness.]{\includegraphics[width = 2in]{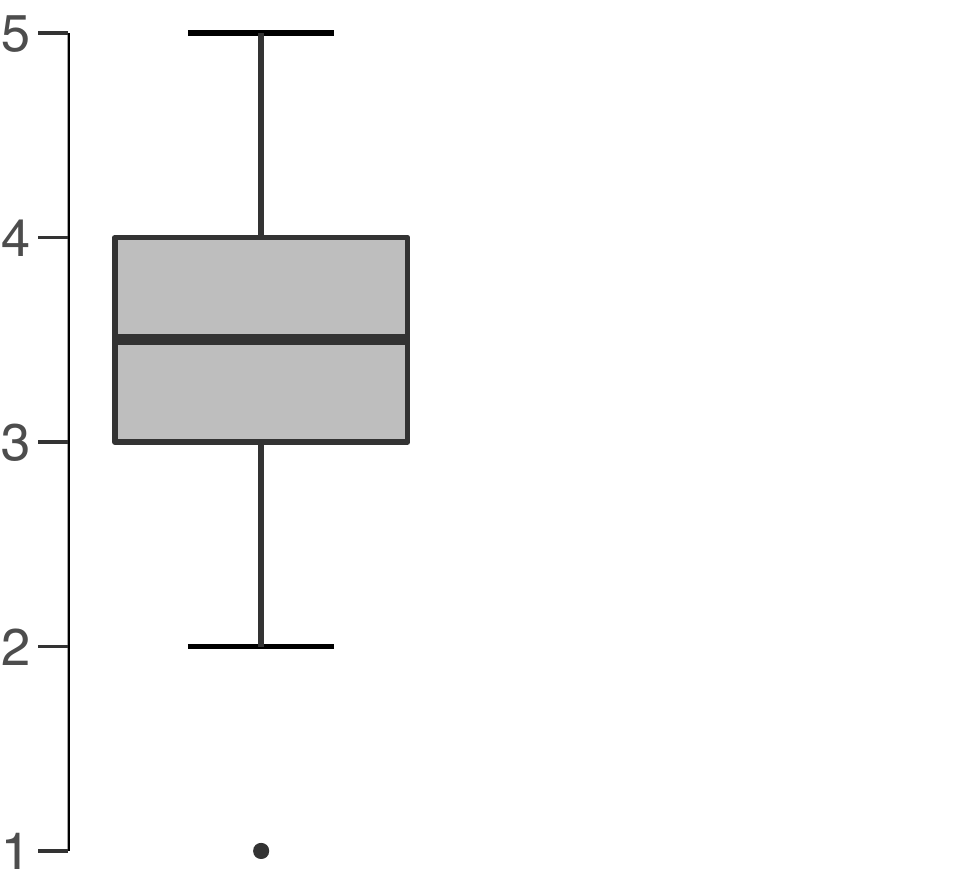}}
\subfloat[Extraversion.]{\includegraphics[width = 2in]{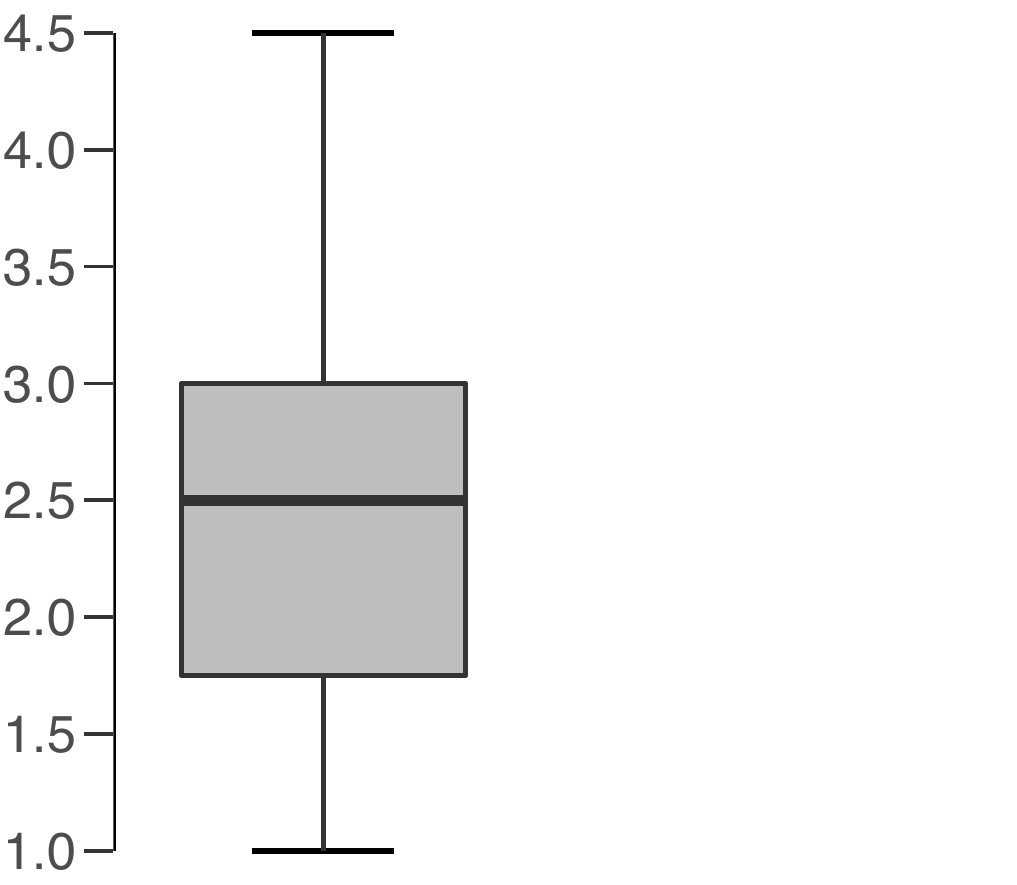}}\\
\subfloat[Agreeableness.]{\includegraphics[width = 2in]{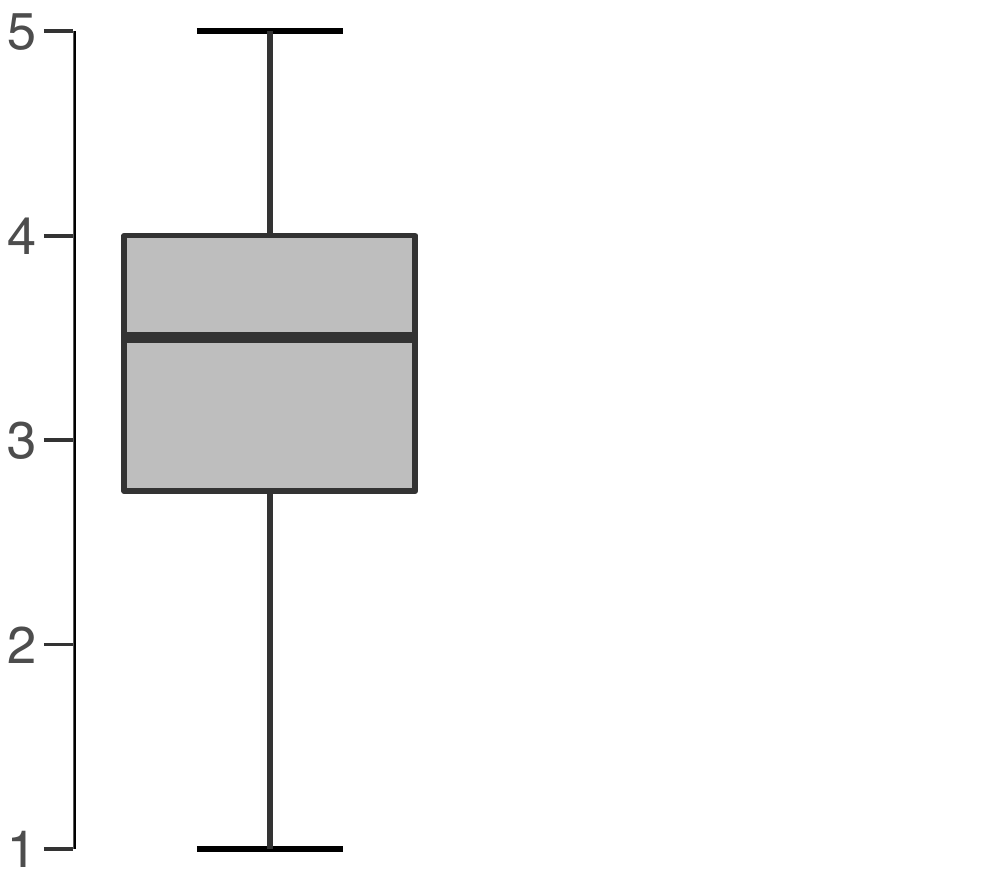}}
\subfloat[Neuroticism.]{\includegraphics[width = 2in]{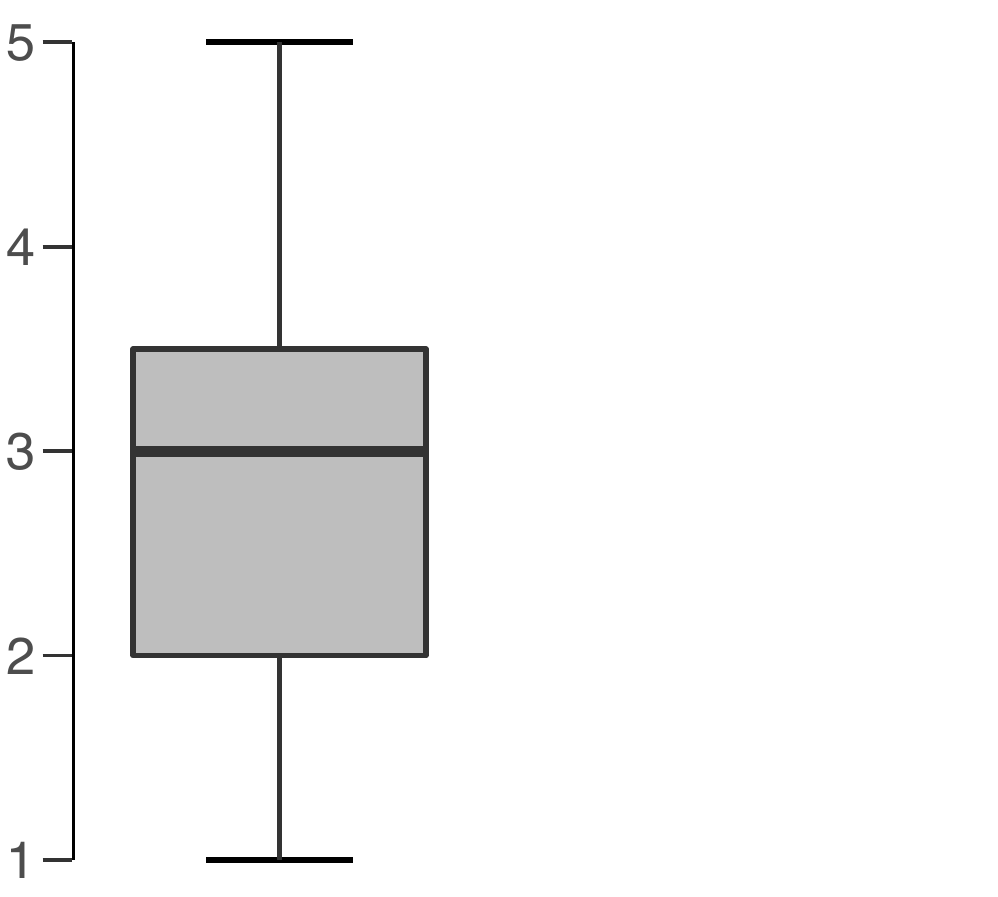}}
\subfloat[Burnout.]{\includegraphics[width = 2in]{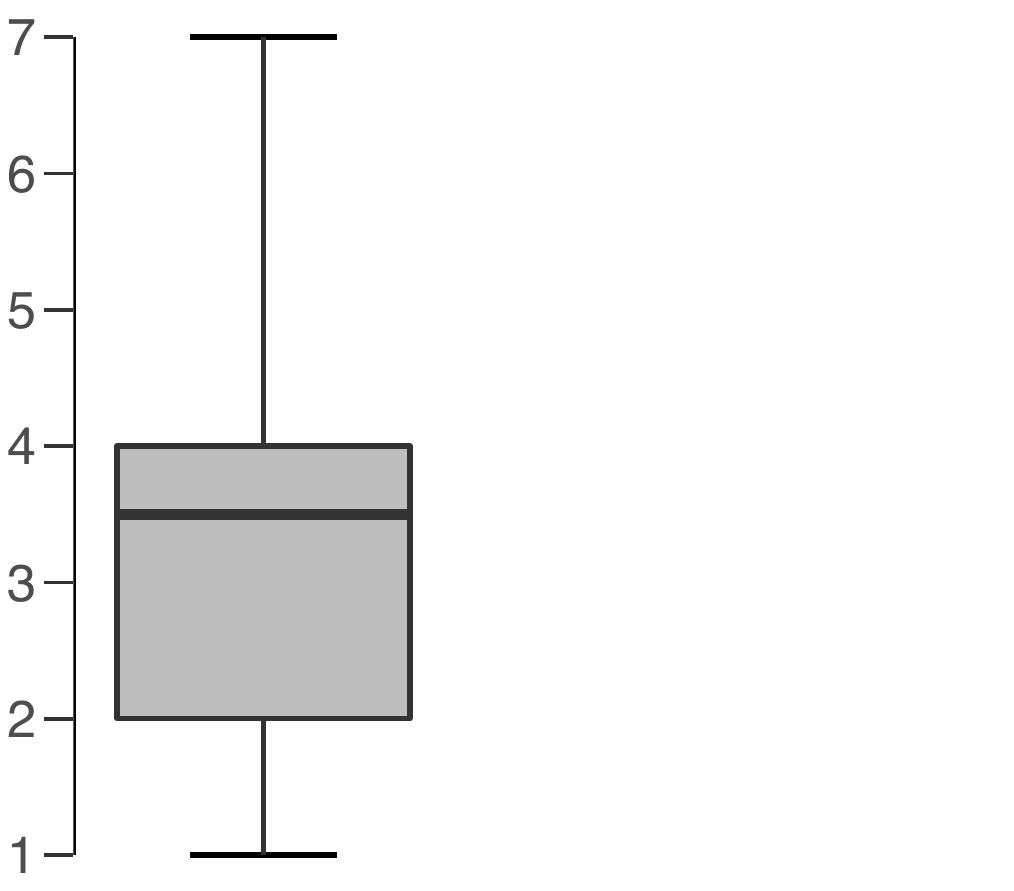}}
\caption{Boxplots of all the constructs in this study.}
\label{boxplots}
\end{figure*}

Figure~\ref{boxplots} shows boxplots for all the six constructs.

The results of the analysis (Table 1) show that the model with only Neuroticism as a predictor outperforms all the other models, confirming the findings of other similar studies in other fields. The other four FFM coefficients are not supported since they do not add any predictive power. Table 1 only shows a few examples as the second model outperformed all the other ones.

\begin{table}[t]\label{table:results}
\small
\caption{Models, Probability of model, Probability of model given data, Bayes Factor 10, and error \% with Burnout as the response variable.}

\begin{tabular}{lcccc}
\toprule
\multicolumn{1}{@{\hspace{2em}}l}{Models} & \mc{$P(M)$} & \mc{$P(M|data)$} & \mc{$BF_{10}$} & \mc{error \%}\\ 
\midrule
null model &                             0.031 & 0.004 & 1 &     \\
Neuroticism &                         0.031 & 0.251 & 60.211 &   0.003 \\
Neuroticism $+$ \\Extraversion &0.031 & 0.112 & 26.892 &   $<$0.001 \\
Neuroticism $+$ \\Conscientiousness &0.031 & 0.111 & 26.606 &   $<$0.001 \\
\midrule\\[-2.5ex]
\multicolumn{3}{l}{} \\
\end{tabular}
\end{table}

\section{Discussion}
The result of our analysis shows that Neuroticism is strongly positively associated with burnout in software developers. This is in line with what has been found by similar studies in other fields \citep{ffm_meta}, but no other personality traits added to the predictive power of the model.

Contrary to the findings of some similar papers on this topic, the result of our study did not find Conscientiousness to have a strong negative association with burnout. This is possibly caused by the small sample size of this study, considering the relatively consistent results in other studies \citep{bigfivefactorspredicators}. An alternative hypothesis is that software development may be different in some significant way to other fields taken as a whole. While Conscientiousness has been shown to be a net positive in resistance to burnout, it has also been shown to be positively correlated with emotional exhaustion \citep{bigfivefactorspredicators}, which contributes to the development of burnout. These somewhat contradicting findings are quite difficult to make sense of without breaking down this constructs even further. This factor may be more pronounced in the field of software development where large ongoing projects and close collaboration are emphasized, but this remains to be explored. 

We would also highlight that did not investigate the quality of work in connection to personality, nor did we take any other confounding factors into account like, for example, teamwork. Therefore, we do not know anything about what effect individuals that score high on neuroticism have on, for example, the quality of the product. We did also not investigate introversion (the opposite of extraversion) in this study, but there seem to be more complex relationship between anxiety, introversion, neuroticism and burnout \citep{williams1992dispositional}. %Or maybe the Bug Five are not such a good reflection of human personality traits and more resolution is needed to understand human behavior in relation to burnout. 

%Här
\subsection{Practical Implications}
Our study has found that software developers scoring high in Neuroticism also experience higher levels of burnout. Companies could be tempted, as an example, make use of this information by testing employee's personalities, both new and current, and using that information to regulate the frequency of the application of a burnout measure, such as SMBM. By testing people with a higher risk of burnout more frequently, less time can be spent on lower risk employees, allowing the discovery of latent burnout faster. In such cases, stress intervention, a form of social support, may be applied. It has been demonstrated to be a strong buffer against the effect of burnout \citep{cobb1976social}. However, we urge careful consideration when using Neuroticism in the employment process as susceptibility to burnout is not exclusively decided by Neuroticism and may be mitigated by other factors, such as decision latitude at work \citep{karasek1979job}. Personality test also do not take any social aspects of the workplace into account, which surely will affect burnout in software developers especially since most of them work on close-knit agile teams.

Lastly, the absence of a strong association with Conscientiousness, despite the consistent results in similar studies, is an interesting finding. While this may be a statistical issue, we speculate that there may be factors specific to the field of software development behind this, and we believe a more detailed investigation into this would give some interesting insight into the specific psychological factors which may be at work within software development. This could, of course also be a validity issue in the used measures.

\section{Conclusion and Future Work}
To our knowledge, this is the first study that has researched the connection between the FFM personality types and burnout in the field of software development. The goal of this study was to analyze the connections between the five different personality types (namely Openness, Conscientiousness, Extraversion, Agreeableness, and Neuroticism) and burnout to see if the result differ from other domains. This study suggests that that Neuroticism is a very strong indicator of workplace burnout, thus confirming previous studies in other domains.

Future work should try to mitigate the selection bias and target software developers, for example, at work. Further studies should also be conducted on larger sample sizes, and with focus on why Conscientiousness might be differently associated to burnout in software developers. Studies could also block confounding factors such as social support or work-family conflict in relation to burnout to get higher resolution of the true effects of personality, however, such studies need a lot of resources, of course.

\bibliographystyle{model5-names}
\bibliography{references}
\newpage
\appendix

\subsection{Mailing Lists}

\begin{itemize}
\footnotesize{
    \item
    Apache Open Office, Development Mailing List:
    \url{https://openoffice.apache.org/mailing-lists.html#development-mailing-list-public}
    \item
    KDE Development:
    \url{https://www.kde.org/support/mailinglists/}
    \item
    KDE Core Development:
    \url{https://www.kde.org/support/mailinglists/}
    \item
    Scilab Developers mailing list: 
    \url{https://www.scilab.org/development/ml}
    \item
    Redhatm software factory-dev:
    \url{https://www.redhat.com/mailman/listinfo/softwarefactory-dev}
    \item
    R-devel: 
    \url{https://www.r-project.org/mail.html}
    \item
    GCC mailing lists:
    \url{https://gcc.gnu.org/lists.html}
    \item
    XMPP for developers: 
    \url{https://xmpp.org/community/mailing-lists.html}
    \item
    Wireshark-dev:
    \url{https://www.wireshark.org/lists/}
    \item
    Django-developers: 
    \url{https://docs.djangoproject.com/en/dev/internals/mailing-lists/}
    \item
    Python.dev:
    \url{https://www.python.org/community/lists/}
    \item
    GNOME, deval-announce-list:
    \url{https://mail.gnome.org/mailman/listinfo/devel-announce-list}
    \item
    TIZEN dev: 
    \url{https://www.tizen.org/community/mailing-lists}
    \item
    VirtualBox developers list: 
    \url{https://www.virtualbox.org/wiki/Mailing_lists}
    \item
    Eclipse Mailing Lists: 
    \url{https://accounts.eclipse.org/mailing-list}
    \item
    Mozilla Web development, General development and Extension development lists: 
    \url{https://www.mozilla.org/en-US/about/forums/#web-development}
    \item
    Debian developers mailing lists:
    \url{https://lists.debian.org/devel.html}
}
\end{itemize}

% Survey in Appendix
\subsection{Survey}

\textbf{Note:} The data from the survey is provided only in aggregate, as per the introductory text in the survey.\\

\noindent\textbf{Validation Section.}
\begin{enumerate}
\item {Are you currently employed as a software engineer?\\  
 (Yes / No)}
\item {How many years of professional software development experience do you have?\\
(None, 1 year, 2-4 years, 5-9 years, 10+ years)}
\end{enumerate}
\textbf{IPIP Section}. Items have the following answer options:
\\ Very Inaccurate | 1, 2, 3, 4, 5 | Very Accurate
\begin{enumerate}
\item I am the life of the party
\item I sympathize with others' feelings	
\item I get chores done right away	
\item I have frequent mood swings	
\item I have a vivid imagination	
\item I don't talk a lot	
\item I am not interested in other people's problems	
\item I often forget to put things back in their proper place	
\item I am relaxed most of the time	
\item I am not interested in abstract ideas	
\item I talk to a lot of different people at parties	
\item I feel others' emotions	
\item I like order	
\item I get upset easily	
\item I have difficulty understanding abstract ideas	
\item I keep in the background	
\item I am not really interested in others	
\item I make a mess of things	
\item I seldom feel blue	
\item I do not have a good imagination	
\end{enumerate}
\textbf{SMBM Section}. Items have the following answer options:
\\ Never/Almost Never | 1, 2, 3, 4, 5, 6, 7 | Always/Almost Always
\begin{enumerate}
\item I feel tired	
\item I have no energy for going to work in the morning	
\item I feel physically drained	
\item I feel fed up	
\item I feel like my "batteries" are "dead"	
\item I feel burned out	
\item My thinking process is slow	
\item I have difficulty concentrating	
\item I feel I'm not thinking clearly	
\item I feel I'm not focused in my thinking	
\item I have difficulty thinking about complex things	
\item I feel I am unable to be sensitive to the needs of coworkers and customers	
\item I feel I am not capable of investing emotionally in coworkers and customers	
\item I feel I am not capable of being sympathetic to coworkers and customers	

\end{enumerate}

\end{document}